\documentclass[conference]{IEEEtran}

\ifCLASSINFOpdf
 \usepackage[pdftex]{graphicx}
  
\else
  
 \usepackage[dvips]{graphicx}
  
\fi

\usepackage[cmex10]{amsmath}
\usepackage{array}
\usepackage{fontenc}
\usepackage{caption}
\usepackage{makeidx}

\usepackage[font=small]{caption}

\usepackage{listings}
\usepackage{fancyhdr}

\fancypagestyle{firstpage}{
  \fancyhf{}
  \fancyfoot[C]{SECURWARE 2013 : The Seventh International Conference on Emerging Security Information, Systems and Technologies}
}
\begin{document}
\pagestyle{fancy}
%
\title{The All-Seeing Eye: A Massive-Multi-Sensor Zero-Configuration Intrusion Detection System for Web Applications}

\author{\IEEEauthorblockN{Christoph Pohl}\\
\IEEEauthorblockA{Munich IT Security Research Group (MuSe)\\
Department of Computer Science and Mathematics\\
Munich University of Applied Sciences\\
Munich, Germany\\
Email: christoph.pohl0@hm.edu}
\and
\IEEEauthorblockN{Hans-Joachim Hof}\\
\IEEEauthorblockA{Munich IT Security Research Group (MuSe)\\
Department of Computer Science and Mathematics\\
Munich University of Applied Sciences\\
Munich, Germany\\
Email: hof@hm.edu}

}

\maketitle
\thispagestyle{firstpage}

\begin{abstract}
\boldmath
Timing attacks are a challenge for current intrusion detection solutions. 
Timing attacks are dangerous for web applications because they may leak information about side channel vulnerabilities. 
This paper presents a massive-multi-sensor zero-configuration Intrusion Detection System that is especially good at detecting timing attacks. 
Unlike current solutions, the proposed Intrusion Detection System uses a huge number of sensors for attack detection. 
These sensors include sensors automatically inserted into web application or into the frameworks used to build web applications. 
With this approach the Intrusion Detection System is able to detect sophisticated attacks like timing attacks or other brute-force attacks with increased accuracy.
The proposed massive-multi-sensor zero-configuration intrusion detection system does not need specific knowledge about the system to protect, hence it offers zero-configuration capability.
\end{abstract}


\section{Introduction}
Intrusion Detection Systems (IDS) in combination with firewalls are the last defense line in security when protecting web applications. The purpose of an IDS is to alert a human operator or a Intrusion Prevention System that an attack is in preparation or currently taking place. 
The configuration of an IDS typically involves a configuration that must be adapted for each system to protect. 
The massive-multi-sensor zero-configuration intrusion detection system for web applications presented in this paper (called All-Seeing Eye in the following) does not need any adaption to the system to protect, hence is very easy to use.

One common challenge for web applications is the detection of timing attacks. A timing attack is an attack, which uses time differences between different actions to gain informations.
Intrusion Detection Systems typically use sensors to collect data. In this work, a sensor describes a data source that provides data useful for attack detection. Useful in this context means that the data must be linked to actions of a web application. Data of sensors is analyzed by All-Seeing Eye to detect attacks. 
Current Intrusion Detection Systems are fairly limited in the number of sensors they use. All-Seeing Eye increases the number of available sensors by injecting code into common web application frameworks or even into web applications to provide additional sensors. A further increase in the number of sensors results from the combination of multiple sensors into one new sensor.
Sensors may collect data from any data source available on a web server, or on the network. Sensors include:
\begin{itemize}
\item Network sensors like TCP requests, or TCP flags (e.g., SYN or SYN/ACK) per period. 
\item Hardware sensors like CPU usage, memory usage, or fan speed.
\item Kernel sensors like number of file IO handles, number of system calls, and the like.
\item Software sensors like log file entries or software hooks.
\item System sensors like alerts from other common Intrusion Detections Systems, e.g., Snort \cite{snort_2013} .
\end{itemize}

The rest of this paper is structured as follows: Section \ref{relwork} presents related work in intrusion detection using multiple sensors. Section \ref{sensors} describes the concept and implementation of the sensors used by All-Seeing Eye. The use of multiple sensors to detect intrusions is described in Section \ref{mmso}. Section \ref{evaluation} evaluates All-Seeing Eye under different attacks, especially timing attacks. Section \ref{conclusion} concludes the paper and gives an outlook on future work.

\section{Related Work}\label{relwork}

Anomaly detection is based on the hypothesis that there are deviations between normal behaviour and behaviour under intrusion \cite{patchapark2007,kruegel_anomaly_2003,liepens_intrusion_1992,denning_intrusion-detection_1987}. 
Many techniques have been researched for the detection like network traffic analysis \cite{lakhina_structural_2004,barford_signal_2002,silveira_urca:_2010}, statistical analysis in records \cite{javitz_sri_1991} or sequence analysis with system calls \cite{lee_data_1998,hofmeyr1998intrusion,forrest_sense_1996,frossi_selecting_2009}.
A combination of this research with anomaly detection methods based on multiple sensors allows to find yet unknown attacks. Configuring intrusion detecting systems for one distinct system or one distinct vulnerability needs configuration with current solutions. The solution presented in this paper does not need any configuration.

In \cite{hofmeyr1998intrusion,forrest_sense_1996}, it is proved that call chains of system calls show different behavior under normal conditions and under intrusion, hence intrusion detection is possible. However, a normal model must be trained using learning data to detect attacks. In \cite{forrest_sense_1996}, it is shown that normal behaviour produces fingerprintable signatures in system call data. 
A deviation from these signatures is defined as intrusion. This method is restricted to the usage of system calls and does not use more fine granular sensor data. In \cite{masri_application-based_2008}, a way to detect anomalies with information flow analysis is shown. Profiling techniques are used, injecting small sensors in a running application. They propose a model with clusters of allowed information flows and compare this normal model against actual information flow. Similar models are proposed in \cite{feng_predicting_2004,bhatkar_dataflow_2006,qin_lift:_2006}. This approach is similar to our approach, but \cite{masri_application-based_2008} focuses on offline audits for penetration testing. The approach presented in this paper is intended to be used online, hence it does not analyze the whole information flow but focuses on the method call chain, and is therefore more efficient.

In \cite{kruegel_multi-model_2005}, it is shown that vulnerability probing can be detected using multiple sensors, especially sensor that calculate the possibility a resource is called by a user. These sensors are called access frequency based sensors. However, the system presented in \cite{kruegel_multi-model_2005} needs a lot of information about the system to protect (e.g., patterns describing legitimate resource calls), hence is difficult to deploy in the field. The solution presented in this paper does not need any configuration.

\section{Sensors for a Massive Multi-Sensor Zero-Configuration Intrusion Detection System}\label{sensors}
\begin{figure}
 \centering
  \includegraphics[width=3.27in]{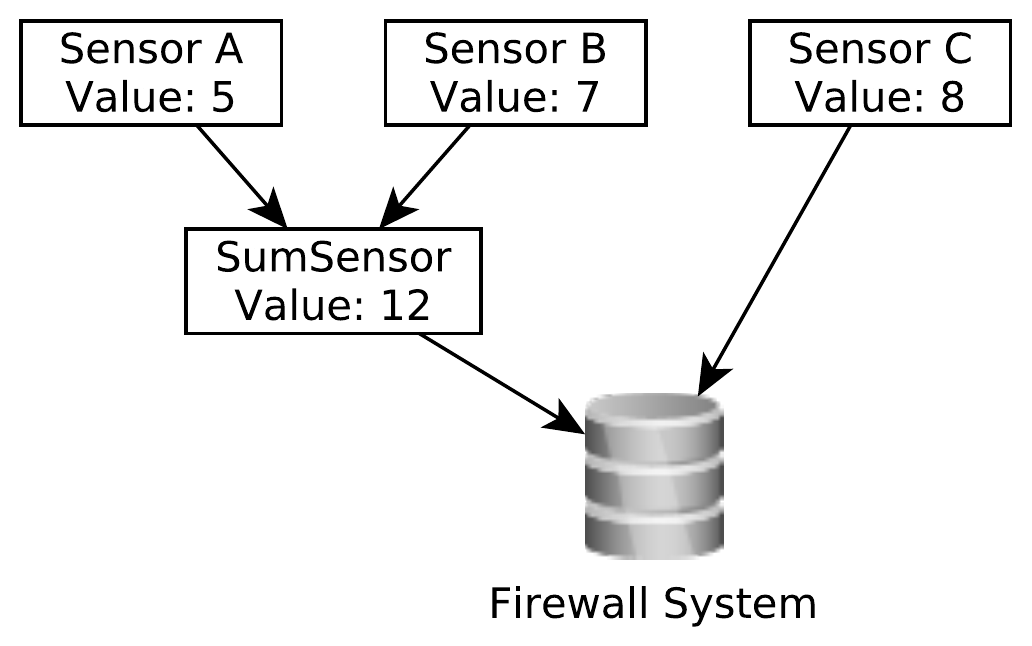}
  
  \caption{Datasource sensors and aggregation sensors}\label{figure:sensors}
\end{figure}

A sensor describes a data source that provides useful data for attack detection. Useful in this context means that the data must be linked to actions of a web application. Data of sensors is analyzed by the proposed Intrusion Detection System All-Seeing Eye to detect attacks. Sensors are:
\begin{itemize}
\item Already available data sources like memory consumption of an application.
\item Software sensors inserted into a web application or a web application framework.
\item Advanced sensors that combine data from multiple other sensors. Figure \ref{figure:sensors}, shows an example of an advanced sensor: sensor A and sensor B are combined into sensor SumSensor by adding the output of sensor A and sensor B at distinct points in time.
\end{itemize}

All-Seeing Eye depends on the availability of a large number of sensors that can be used for attack detection. The current implementation of All-Seeing Eye is using the following classes of sensors:
\begin{itemize}
	\item Alert Sensors: alert sensors have an upper and/or a lower boundary for data measured by this sensor that triggers an alert.
	\item Sensors with non-string payload: 
	\begin{itemize}
	\item 64 Bit payload: These sensors have a payload. The payload typically consists of measurement values (e.g., time differences).
	\item 32 Bit payload: The payload consists of four byte and is typically used for simple state (e.g. the user authentication state, either "not authenticated" or "authenticated" ).
	\end{itemize} 
	\item String Sensor: the payload of sensors of this class consists of a simple string. This class of sensors is usually used in regular expressions.
	\item Filter Sensor: a sensor of this class filters the output of an existing sensor, e.g. by evaluating regular expressions.
	\item Aggregation Sensors: sensors of this class combine output of other sensors into a new sensor. Based on the time of aggregation, two subclasses are used:
	\begin{itemize}
	\item One-Time Aggregation Sensor: a sensor of this class can combine values of other sensors, e.g. by addition, multiplication or logarithmic scaling.
	\item Continuous Aggregation Sensor: sensors of this class combine values of other sensors over a certain period of time, creating an average or quantile.	
	\end{itemize}
\end{itemize}

The current implementation of All-Seeing Eye uses software sensors, but the presented approach can be extended to hardware sensors. However, hardware sensors are not within the scope of this paper. All-Seeing Eye is intended to protect web applications relying on the Java runtime, however, the massive-multi-sensor zero-configuration approach may also be adapted to other execution environments. In the following, it is described how software sensors can be realized.

\subsection{Sensor Implementation}
Software sensors are implemented by injecting hooks at the beginning and end of methods of a web application. Hence, hooks are called before and after code execution of a method. With this approach, it is e.g. possible to measure the method execution time for each method used. It is also possible to identify the order of method execution. Hooks are injected directly into Java bytecode. It is not necessary to recompile any Java web application protected by All-Seeing Eye. Deploying All-Seeing Eye is as simple as copying the All-Seeing Eye jar file into the library directory of the web server. It is not necessary to perform any configuration for the web application that should be protected, hence All-Seeing Eye is called "zero configuration". As injection technology AspectJ \cite{aspectj_2013}, with Load Time Weaving \cite{weaving_2013}, is used. 
For the testbed used for the evaluation presented in this paper the aspect is placed in OpenCMS \cite{opencms_2013}. OpenCMS is a well known and widely used framework for Content Management.
All-Seeing Eye takes care that methods used by the protected web application do not clash with method names used by All-Seeing Eye. Sensor data is written to a log file for further analyses.

A typical software sensor will produce data as followed:

\begin{lstlisting}
timestamp,count,SID,value
\end{lstlisting}

where $Timestamp$ and $count$ together form a unique ID. $Timestamp$ is a UNIX timestamp of the event. $Count$ is a counter incremented each time adding a SID to a timestamp with at least one SID at the same timestamp. $Values$ is the sensor value. $SID$ is a unique key describing one sensor in this format:

\begin{lstlisting}
package.class.method.id.vid.vvid
\end{lstlisting}

where $package$ is the package in which the $class$ is located in which the $method$ is situated that was injected for this sensor. $Id$ is used to distinguish overloaded methods from each other. $Vid$ identifies a type of the value (e.g., String), and $vvid$ is an additional sensor for further use (e.g., if two sensors are injected into the same method).

A software sensor is unique with 
\begin{lstlisting}
package.class.method.id.vid.vvid
\end{lstlisting}

One way to minimize the output of sensors (and the number of data to write to the log file) is to produce no output for methods that have an execution time lower than the resolution of the timestamps (1 ms). It is suspected that these methods would not generate any interesting output as these methods are usually helper methods or wrappers.

\subsection{Memory Consumption and Computational Overhead of the Proposed Multi-Sensor Approach}
Software sensors have an impact on memory consumption of the protected web application. An additional 30 byte code operation are added to each method call of the web application by All-Seeing Eyes. Hence, the percentage of computational overhead heavily depends on the length of the methods of the web application. 

\begin{figure}
 \centering
\includegraphics[width=3.27in]{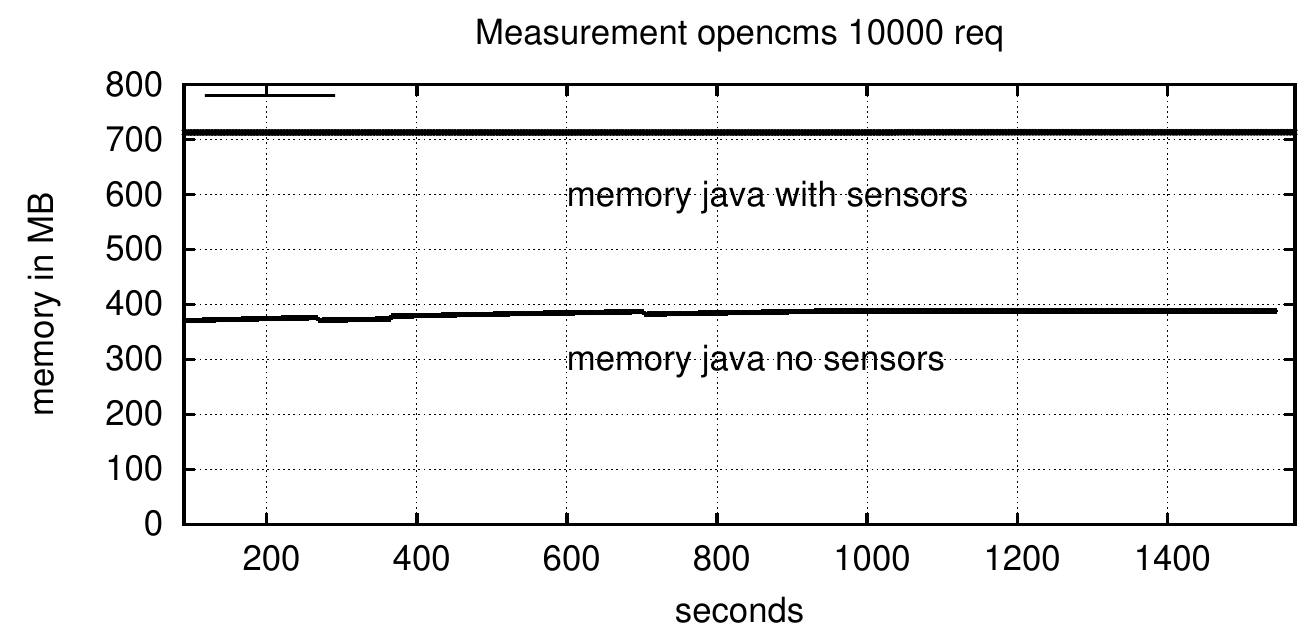} 
\caption{Memory comsumption of All-Seeing Eye}\label{figure:memorylog-crop}
\end{figure}

In the testbed using OpenCMS (see Section \ref{evaluation}), the following experiment was conducted to evaluate the memory and computation overhead:

10,000 requests are sent to a web application using All-Seeing Eye. 100 runs were used, resetting the server after each run. Figure \ref{figure:memorylog-crop} shows the average memory consumption from 100 runs of the experiment compared to the memory consumption from 100 runs of the web application without All-Seeing Eye.
 It can be seen that a significant overhead of approximately 300 MB is needed to run All-Seeing Eye. This overhead is the result of buffering data before writing to the log files.

\section{Massive Multi-Sensor Zero-Configuration Intrusion Detection System}\label{mmso}
This Section describes the design of the proposed massive multi-sensor zero-configuration intrusion detection system All-Seeing Eye.
All-Seeing Eye uses the software sensors, described in more detail in the last section, to calculate intrusion metrics. 
The metrics described in the following are focused on detection of outliers in timeline data values to detect brute force attacks.
However, the approach presented in this paper is not limited to this attack class, it can be easily adapted to detect various other attacks.
Even attacks on the business logic can be detected as the presented approach uses software sensors embedded in the code of an application. This is out of scope of this paper.

An advantage of All-Seeing Eye is that it allows to detect side channel attacks without knowledge of the web application which is to be protected. 
In the absence of an attack, there is  a high correlation between method calls defined in a method chain. As shown in Section \ref{evaluation} a single call results in correlated calls (method chain) of other methods.
 The system under load shows the same correlations. These correlations are further called as fingerprint $s$. 
 Under attack, however, the system shows a different behavior, hence allows to identify attacks, see Section \ref{evaluation} for details.
All-Seeing Eye does not need a preconfigured or constructed normal model. For this approach the normal model is created from history.
At time $t=0$ it is always assumed that there is no attack, hence status $c$ is always $c!=attack$. 
If there is no attack, the same fingerprint $s$ should show up in each distinct time period $T$ with the same probability.
A deviation from the number of fingerprings (written as $|s|$) in a time period $T$ is defined as possible intrusion.
This behaviour is well known, as stated in Section \ref{relwork}. The new approach here is the lack of need to define what a similar request is.
The normal model is built using a quantile function, where the result is called  $\alpha$.
$\alpha$ uses a floating history time period, which is defined as $n \times T$ and $t \in T$ are in state $c!=attack$.The multiplier $n$ defines how much of the history is used. To control the sensitivity of the system, a configuration parameter $p$ is used.
In normal model $\alpha$, a deviation is detected by:
\begin{equation}
c=\begin{cases}
attack & \text{if } |s_{currentT}| \geq \alpha \times p\\
!attack & \text{if } |s_{currentT}| < \alpha \times p
\end{cases}
\end{equation}
This calculation is robust against statistical outliers and can be evaluated fast enough for real time calculation, in combination with structures related to sort optimisation. In further researches these calculations will be done (together with other sensor calculations) with a graphical processing unit.


\section{Evaluation}\label{evaluation}
For the evaluation of the massive multi-sensor zero-configuration intrusion detection system, two typical attacks on web applications are used: timing attacks and vulnerability probing. Especially timing attacks are hard to detect for common intrusion detection systems.
For our test environment OpenCMS version 8.5.1 \cite{opencms_2013} is used as web application to protect. OpenCMS is a well known and widely used framework for Content Management. 

\subsection{Evaluation Environment}
For the evaluation of All-Seeing Eye, a paravirtualized, openvz solution \cite{openvz_2013}  is used. 
This approach has the advantage that is is very realistic compared to simulations. 
The presented hardware settings are the settings of the corresponding virtual machine. 
Table \ref{tab:metcond} lists hardware and software used for the evaluation.

\begin{table}[htpb]
\centering
\caption{Experimental setup}\label{tab:metcond}
    \begin{tabular}{|c|c|}  
\hline
	\hline
\multicolumn{2}{|c|}{\textbf{Hardware(Server)}}\\
\hline
CPU & 4 Cores (2.1GHZ on hostsystem)\\
Memory & 6 GB Ram\\
Ethernet & Bridged at 1 GBit Nic \\
\hline
\multicolumn{2}{|c|}{\textbf{Software(Server)}}\\
\hline
Server Version & Apache Tomcat 7.0.28 \cite{tomcat_2013}\\
JVM & Sun 1.6.0.27-b27\\
    \hline
    \end{tabular}
    
\end{table}

\subsection{Fingerprints of Normal Behaviour}
To validate the hypothesis, that requests to the same target have the same fingerprint, the following experiment has been conducted.

First, a baseline is established for all other experiments. To do so, several requests are sent to the server and the server is restarted after each request. No interfering processes are running on this server.

\begin{figure}
 \centering
\includegraphics[width=3.27in]{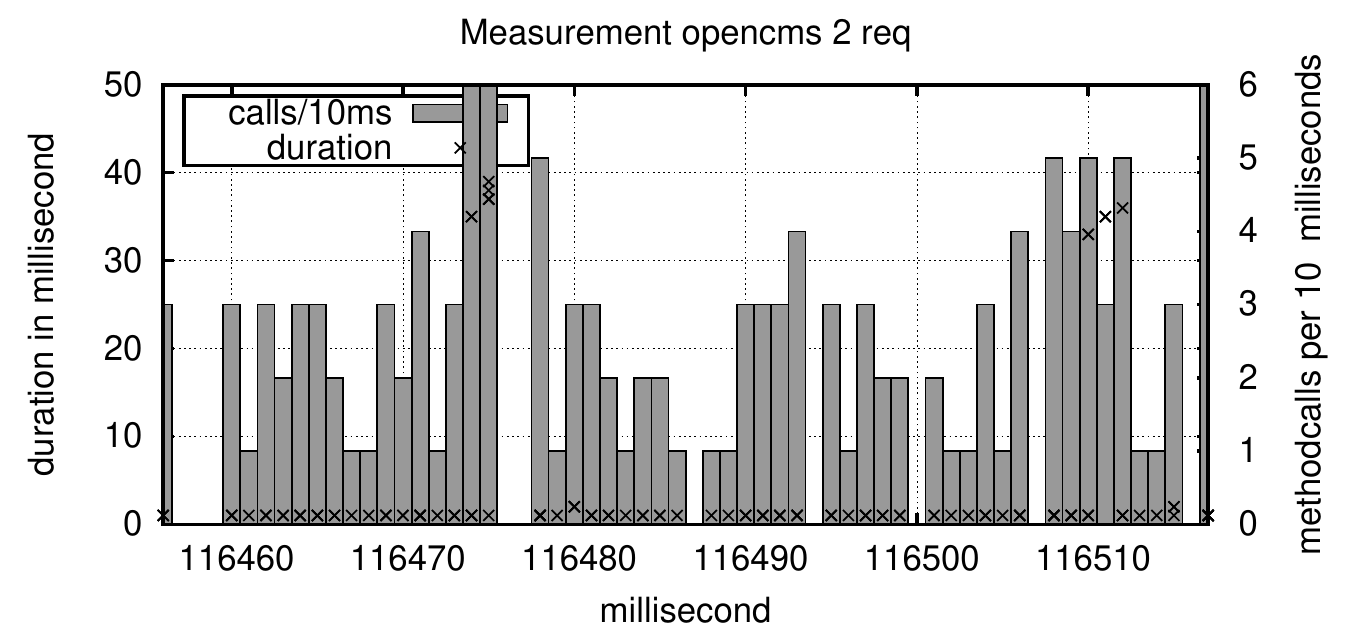} 
\caption{Two fingerprints of different requests}\label{figure:twodifferentprints}
\end{figure}

\begin{figure}
 \centering
\includegraphics[width=3.27in]{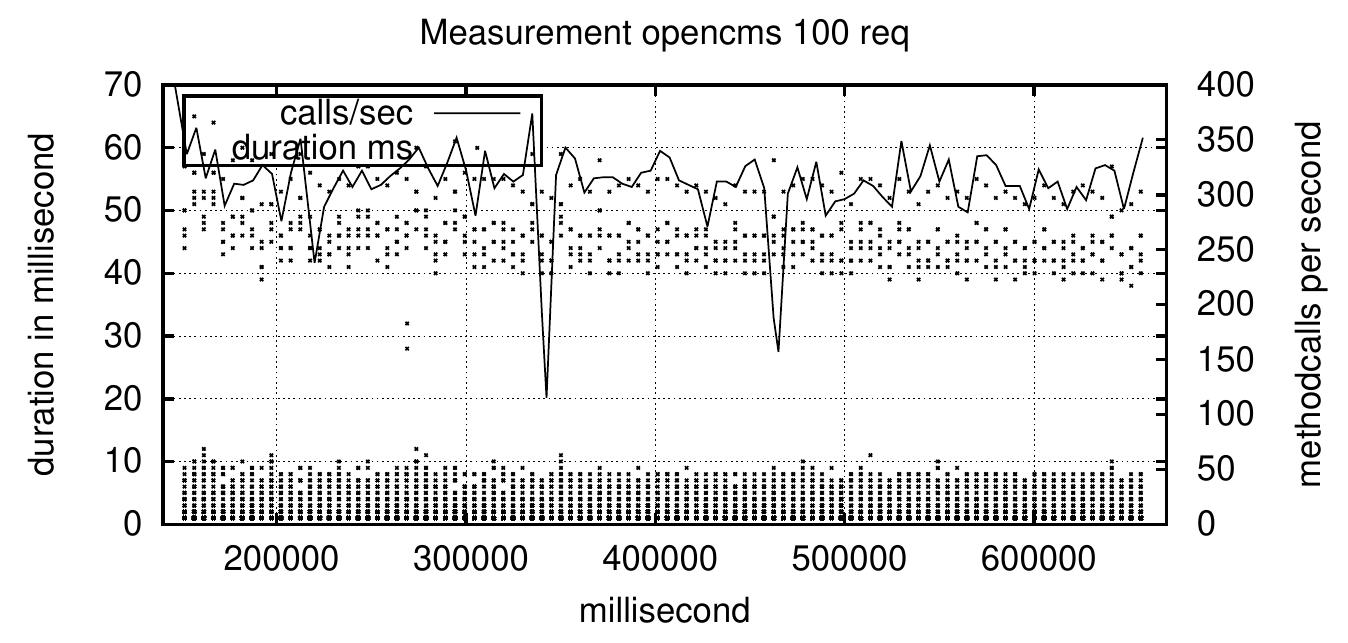}
\caption{100 requests on the same page}\label{figure:trace100}
\end{figure}

After establishing the baseline, the whole website is crawled in a second step, ensuring that requests are sequential. The crawler is configured to request a single page and all depending images and scripts. To test the software under load in the third step, another crawler requests the server with 20 concurrent users, with a delay of one second between each request/user. Overall there were in average 20 requests per second for different websites. The second and third step have been repeated 100 times. In each run of the experiment, the metrics described above produced unique fingerprints for every requested target. This can be seen in Figure \ref{figure:twodifferentprints}.
In this figure the fingerprint of the start page and the request to the login page are extracted from the logged data.
Under load the signature looks like the picture presented in Figure \ref{figure:trace100}.

The result in Figure \ref{figure:twodifferentprints} shows that there are stable correlations between method calls. For better readability, points that differ less then 3 milliseconds are averaged. The experiments show that it is possible with All-Seeing Eye to identify similar requests using their fingerprint.

\subsection{Fingerprints of Information Leakage and Probing for Vulnerabilities}\label{probing}
OpenCMS version 8.5.1 has a known information leakage vulnerability, as described in \cite{owasp_2013}: using the default setting there is no limit for failed logins per time period. Also, a large amount of information is given in error messages, especially the error message "this username is unknown", if the given user name does not exist and "password is wrong", if the given password is wrong for an existing user allow an attacker to find valid user names, by trying possible user names from a dictionary and using error messages to find out if an user name is valid. This attack can be detected with a statistical analysis to detect the brute force analysis \ref{relwork}.
To do so, a detection technique needs to identify if a single resource is called many times but with different parameter in the request header. A normal model is needed for allowing patterns to test for deviations of the normal model. This needs deep knowledge of the system to protect and the vulnerability itself. All-Seeing Eye is able to detect this attack (and also other probings using brute force attacks), without this knowledge about system and vulnerability.

To evaluate if All-Seeing Eye can recognize brute-force attacks, an experiment has been conducted where an attacker probes the login page and tries to identify valid user names. The following pattern was used to generate the login requests:

\begin{lstlisting}
http://192.168.2.89:8080/opencms/.../index.html?action=login&username=username_1&password=passwordnotindb...snippedEnd
//username_1..username_n in dictionary
http://192.168.2.89:8080/opencms/.../index.html?action=login&username=username_n&password=passwordnotindb...snippedEnd
\end{lstlisting}

The attacker used a dictionary with 1000 names for the brute-force attack. To make detection harder, the attacker uses 50 different user agents as well as 20 different IPs. Only one valid username exists in the database.

Figure \ref{figure:trace100} shows a subset of 100 requests. From the figure it is obvious, that the probing attemps produce many similar fingerprints. It shows a high correlation between different requests, the sensor values and the order different sensors are called in one requests. This order and the values are stable over all requests. Hence, All-Seeing Eyes can easily detect a probing attack even if someone uses different header data. No a-priori knowledge of the system which is to be protected or the vulnerability itself is necessary.

\subsection{Fingerprints of Timing attacks}\label{timing}
\begin{figure}
 \centering
\includegraphics[width=3.27in]{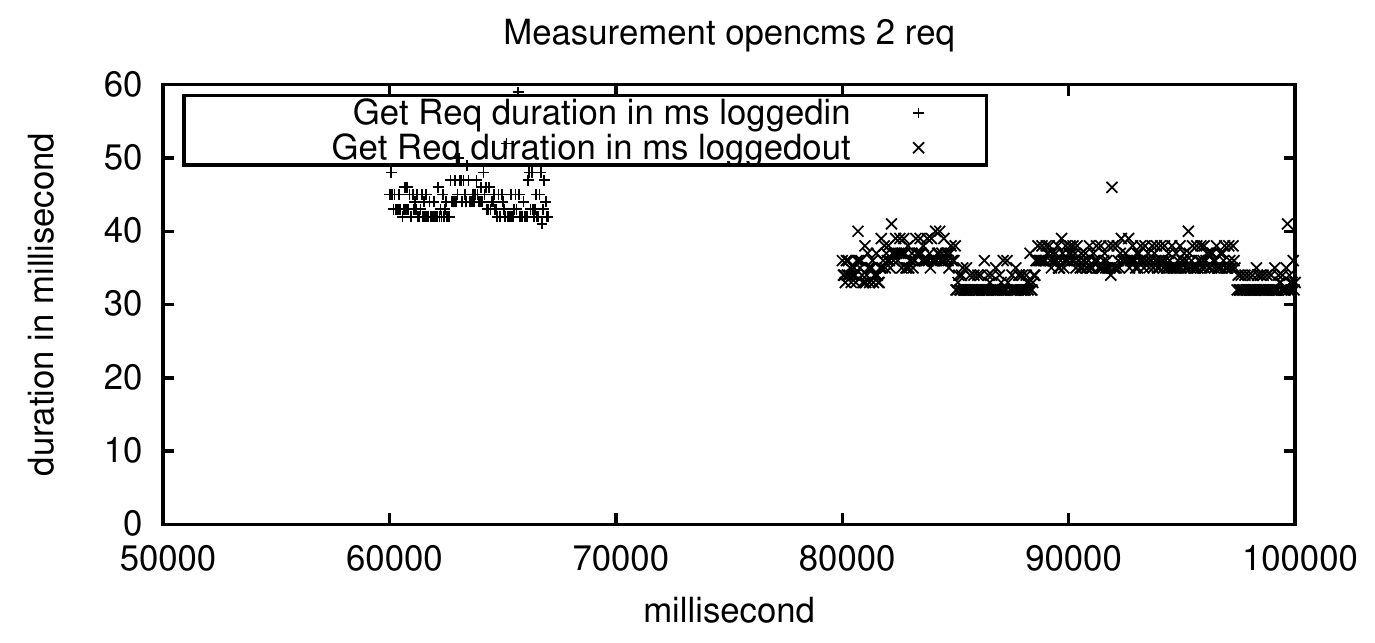}
\caption{Time difference between logged in users and users not logged in on the start page}\label{figure:loggedin}
\end{figure}

OpenCMS version 8.5.1 is vulnerable to timing attacks as can be seen in Figure \ref{figure:loggedin}. The figure shows the times for loading of the start page for users that are already logged in as well as for users that are not logged in.  A significant difference (849 ms to 798 ms) exists.

Using this timing difference an attacker can brute force user names by a dictionary attack. All logged in users can be detected. As with the information leakage and probing attack in Subsection \ref{probing}, current intrusion detection solutions need information about the system which is to be protected and the vulnerability to detect this attack.

To test if All-Seeing Eye is able to detect timing attacks without knowledge (zero-Configuration), the following experiment has been conducted: An attacker uses a dictionary of 1000 user names to execute the timing attack. Each request has different header data in the request only in the login name and the password as shown in following listing:
\begin{lstlisting}
//successful login
http://192.168.2.89:8080/opencms/.../index.html?action=login&username=admin98&password=admin1...snippedEnd
//username not present, pwd not present
http://192.168.2.89:8080/opencms/.../index.html?action=login&username=usernamenotpresent&password=wrongpwd...snippedEnd
//username present with wrong password
http://192.168.2.89:8080/opencms/.../index.html?action=login&username=admin832&password=wrongpwd...snippedEnd
\end{lstlisting}

\begin{figure}
 \centering
\includegraphics[width=3.27in]{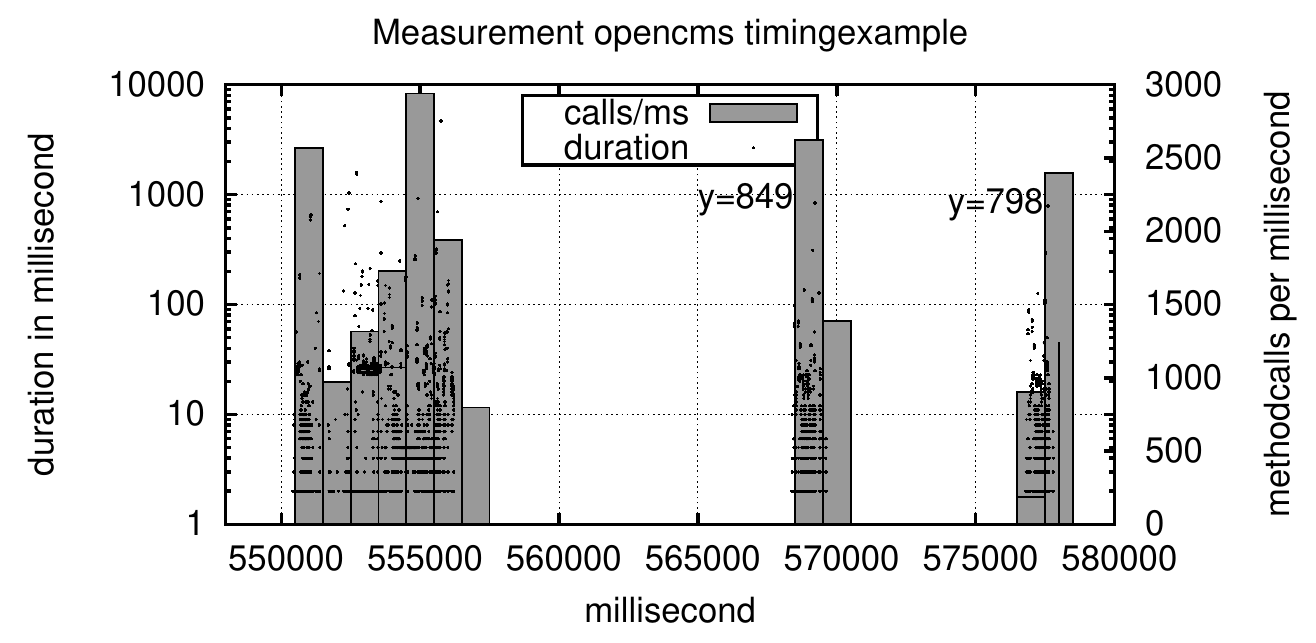}
\caption{100 requests on the same page}\label{figure:traceTiming}
\end{figure}

Figure \ref{figure:traceTiming} shows an examples of a fingerprints of the experiment. The first fingerprint shows a successful login, the second fingerprint shows a login with no present username and third fingerprint shows a login with present username but wrong password. The results clearly show a correlation of the differences in order, time, and amount of method calls for each request. The presence of unusual high amount of similar fingerprints in a distinct time period allows to detect this timing attack. Hence, All-Seeing Eye can identify timing attacks.

\section{Conclusion}\label{conclusion}
This paper presented the massive-multi-sensore zero-configuration intrusion detection solution All-Seeing Eye. All-Seeing Eye uses multiple software sensors for attack detection. Software sensors are automatically injected into web applications to protect. No change to deployed web applications is necessary and All-Seeing Eye does not need to be configured for each web application. All-Seeing Eye is zero-configuration. The evaluation showed that brute force attacks, e.g. timing attacks, can be detected by All-Seeing Eye using fingerprints automatically generated. It also shows that it is possible to identify similar requests without knowledge over the system or the definition of patterns. The current version of All-Seeing Eye produces a memory overhead of approximately 300 MB, which will be reduced in future versions using Graphical Processing Units (GPU) for fast calculation, reducing the need to write data to log files. This will also increase the amount of possible sensors, which can be calulated near real time. Further work will target different attacks like Blind SQL-injection, CSRF attacks, XSS attacks, and Command Injection.

\section*{Acknowledgment}
This work is part of the project ``Sichere Entwicklung und Sicherer Betrieb von Webanwendungen`` of the Munich IT Security Research Group funded by the Bayerisches Staatsministerium f\"ur Wissenschaft, Forschung und Kunst.

\bibliographystyle{IEEEtran}


\begin{thebibliography}{10}
\providecommand{\url}[1]{#1}
\csname url@samestyle\endcsname
\providecommand{\newblock}{\relax}
\providecommand{\bibinfo}[2]{#2}
\providecommand{\BIBentrySTDinterwordspacing}{\spaceskip=0pt\relax}
\providecommand{\BIBentryALTinterwordstretchfactor}{4}
\providecommand{\BIBentryALTinterwordspacing}{\spaceskip=\fontdimen2\font plus
\BIBentryALTinterwordstretchfactor\fontdimen3\font minus
  \fontdimen4\font\relax}
\providecommand{\BIBforeignlanguage}[2]{{%
\expandafter\ifx\csname l@#1\endcsname\relax
\typeout{** WARNING: IEEEtran.bst: No hyphenation pattern has been}%
\typeout{** loaded for the language `#1'. Using the pattern for}%
\typeout{** the default language instead.}%
\else
\language=\csname l@#1\endcsname
\fi
#2}}
\providecommand{\BIBdecl}{\relax}
\BIBdecl

\bibitem{snort_2013}
\BIBentryALTinterwordspacing
``Snort :: Home page,'' http://www.snort.org/, retrieved 2013-04-11. [Online].
  Available: \url{http://www.snort.org/}
\BIBentrySTDinterwordspacing

\bibitem{patchapark2007}
\BIBentryALTinterwordspacing
A.~Patcha and J.~. Park, ``An overview of anomaly detection techniques:
  Existing solutions and latest technological trends,'' Computer
  Networks, vol.~51, no.~12, pp. 3448--3470, 2007
\BIBentrySTDinterwordspacing

\bibitem{kruegel_anomaly_2003}
\BIBentryALTinterwordspacing
C.~Kruegel and G.~Vigna, ``Anomaly detection of web-based attacks,'' in
  Proceedings of the 10th {ACM} conference on Computer and communications
  security, ser. {CCS} '03.\hskip 1em plus 0.5em minus 0.4em\relax New York,
  {NY}, {USA}: {ACM}, 2003, p. 251–261
\BIBentrySTDinterwordspacing

\bibitem{liepens_intrusion_1992}
\BIBentryALTinterwordspacing
G.~Liepens and H.~Vaccaro, ``Intrusion detection: Its role and validation,''
  Computers \& Security, vol.~11, no.~4, pp. 347--355, Jul. 1992
\BIBentrySTDinterwordspacing

\bibitem{denning_intrusion-detection_1987}
D.~Denning, ``An intrusion-detection model,'' {IEEE} Transactions on
  Software Engineering, vol. {SE-13}, no.~2, pp. 222--232, 1987

\bibitem{lakhina_structural_2004}
\BIBentryALTinterwordspacing
A.~Lakhina, K.~Papagiannaki, M.~Crovella, C.~Diot, E.~D. Kolaczyk, and N.~Taft,
  ``Structural analysis of network traffic flows,'' in Proceedings of the
  joint international conference on Measurement and modeling of computer
  systems, ser. {SIGMETRICS} {'04/Performance} '04.\hskip 1em plus 0.5em minus
  0.4em\relax New York, {NY}, {USA}: {ACM}, 2004, p. 61–72
\BIBentrySTDinterwordspacing

\bibitem{barford_signal_2002}
\BIBentryALTinterwordspacing
P.~Barford, J.~Kline, D.~Plonka, and A.~Ron, ``A signal analysis of network
  traffic anomalies,'' in Proceedings of the 2nd {ACM} {SIGCOMM} Workshop
  on Internet measurment, ser. {IMW} '02.\hskip 1em plus 0.5em minus
  0.4em\relax New York, {NY}, {USA}: {ACM}, 2002, p. 71–82
\BIBentrySTDinterwordspacing

\bibitem{silveira_urca:_2010}
F.~Silveira and C.~Diot, ``{URCA:} pulling out anomalies by their root
  causes,'' in 2010 Proceedings {IEEE} {INFOCOM}, 2010, pp. 1--9

\bibitem{javitz_sri_1991}
H.~Javitz and A.~Valdes, ``The {SRI} {IDES} statistical anomaly detector,'' in
   1991 {IEEE} Computer Society Symposium on Research in Security and
  Privacy, 1991. Proceedings, 1991, pp. 316--326

\bibitem{lee_data_1998}
\BIBentryALTinterwordspacing
W.~Lee and S.~J. Stolfo, ``Data mining approaches for intrusion detection,'' in
  Proceedings of the 7th conference on {USENIX} Security Symposium -
  Volume 7, ser. {SSYM'98}.\hskip 1em plus 0.5em minus 0.4em\relax Berkeley,
  {CA}, {USA}: {USENIX} Association, 1998, p. 6–6
\BIBentrySTDinterwordspacing

\bibitem{hofmeyr1998intrusion}
S.~Hofmeyr, S.~Forrest, and A.~Somayaji, ``Intrusion detection using sequences
  of system calls,'' Journal of computer security, vol.~6, no.~3, pp.
  151--180, 1998

\bibitem{forrest_sense_1996}
S.~Forrest, S.~Hofmeyr, A.~Somayaji, and T.~Longstaff, ``A sense of self for
  unix processes,'' in  1996 {IEEE} Symposium on Security and Privacy,
  1996. Proceedings, 1996, pp. 120--128

\bibitem{frossi_selecting_2009}
A.~Frossi, F.~Maggi, G.~L. Rizzo, and S.~Zanero, ``Selecting and improving
  system call models for anomaly detection,'' in Detection of Intrusions
  and Malware, and Vulnerability Assessment, ser. Lecture Notes in Computer
  Science, U.~Flegel and D.~Bruschi, Eds.\hskip 1em plus 0.5em minus
  0.4em\relax Springer Berlin Heidelberg, Jan. 2009, no. 5587, pp. 206--223

\bibitem{masri_application-based_2008}
\BIBentryALTinterwordspacing
W.~Masri and A.~Podgurski, ``Application-based anomaly intrusion detection with
  dynamic information flow analysis,'' Computers \& Security, vol.~27,
  no. 5–6, pp. 176--187, Oct. 2008
\BIBentrySTDinterwordspacing

\bibitem{feng_predicting_2004}
\BIBentryALTinterwordspacing
L.~Feng, X.~Guan, S.~Guo, Y.~Gao, and P.~Liu, ``Predicting the intrusion
  intentions by observing system call sequences,'' Computers \&
  Security, vol.~23, no.~3, pp. 241--252, May 2004
\BIBentrySTDinterwordspacing

\bibitem{bhatkar_dataflow_2006}
S.~Bhatkar, A.~Chaturvedi, and R.~Sekar, ``Dataflow anomaly detection,'' in
  2006 {IEEE} Symposium on Security and Privacy, 2006, pp. 15 pp.--62

\bibitem{qin_lift:_2006}
F.~Qin, C.~Wang, Z.~Li, H.~Kim, Y.~Zhou, and Y.~Wu, ``{LIFT:} a low-overhead
  practical information flow tracking system for detecting security attacks,''
  in 39th Annual {IEEE/ACM} International Symposium on Microarchitecture,
  2006. {MICRO-39}, 2006, pp. 135--148

\bibitem{kruegel_multi-model_2005}
\BIBentryALTinterwordspacing
C.~Kruegel, G.~Vigna, and W.~Robertson, ``A multi-model approach to the
  detection of web-based attacks,'' Computer Networks, vol.~48, no.~5,
  pp. 717--738, Aug. 2005
\BIBentrySTDinterwordspacing

\bibitem{aspectj_2013}
\BIBentryALTinterwordspacing
``{AspectJ} aspectj homepage,'' {http://www.eclipse.org/aspectj}, retrieved
  2013-04-11. [Online]. Available: \url{http://www.eclipse.org/aspectj}
\BIBentrySTDinterwordspacing

\bibitem{weaving_2013}
\BIBentryALTinterwordspacing
``{Load Time Weaving} developer guide aspectj,''
  {http://eclipse.org/aspectj/doc/released/devguide/ltw.html}, retrieved
  2013-04-11. [Online]. Available:
  \url{http://eclipse.org/aspectj/doc/released/devguide/ltw.html}
\BIBentrySTDinterwordspacing

\bibitem{opencms_2013}
\BIBentryALTinterwordspacing
``{OpenCms}, opencms homepage,'' http://www.opencms.org, retrieved 2013-04-11.
  [Online]. Available: \url{http://www.opencms.org}
\BIBentrySTDinterwordspacing

\bibitem{openvz_2013}
\BIBentryALTinterwordspacing
``{OpenVZ} openvz linux containers,'' {http://openvz.org}, retrieved
  2013-04-11. [Online]. Available: \url{http://openvz.org}
\BIBentrySTDinterwordspacing

\bibitem{tomcat_2013}
\BIBentryALTinterwordspacing
``{Apache Tomcat} apache tomcat,'' {http://tomcat.apache.org}, retrieved
  2013-04-11. [Online]. Available: \url{http://tomcat.apache.org}
\BIBentrySTDinterwordspacing

\bibitem{owasp_2013}
\BIBentryALTinterwordspacing
``{Owasp Top Ten},'' {https://www.owasp.org/index.php/Category:
  OWASP\_Top\_Ten\_Project}, retrieved 2013-04-11. [Online]. Available:
  \url{https://www.owasp.org/index.php/Category: OWASP\_Top\_Ten\_Project}
\BIBentrySTDinterwordspacing

\end{thebibliography}

\end{document}